\documentclass[%
aip,
reprint,
superscriptaddress,
ymb,
prb,
twocolumn,
titlepage,
floatfix,
showpacs,
]{revtex4-2}

\usepackage{graphicx}
\usepackage{hyperref}
\usepackage{amssymb}
\usepackage[fleqn]{amsmath}
\usepackage{textcomp}
\usepackage{xcolor}
\allowdisplaybreaks

\definecolor{linkblue}{RGB}{49,49,148}
\hypersetup{linkcolor  = linkblue,citecolor  = linkblue,urlcolor   = linkblue,colorlinks = true,}

\makeatletter
\renewcommand*{\eqref}[1]{%
  \hyperref[{#1}]{\textup{\tagform@{\ref*{#1}}}}%
}
\makeatother

\begin{document}

\title{Non-thermal electrons open the non-equilibrium pathway of the phase transition in FeRh}

\author{M.~Mattern}
\affiliation{Institut f\"ur Physik und Astronomie, Universit\"at Potsdam, 14476 Potsdam, Germany}
\affiliation{Max-Born-Institut f\"ur Nichtlineare Optik und Kurzzeitspektroskopie, 12489 Berlin, Germany}
\author{S. P.~Zeuschner}
\affiliation{Institut f\"ur Physik und Astronomie, Universit\"at Potsdam, 14476 Potsdam, Germany}
\author{M.~R\"ossle}
\affiliation{Helmholtz-Zentrum Berlin f\"ur Materialien und Energie GmbH, Wilhelm-Conrad-R\"ontgen Campus, BESSY II, 12489 Berlin, Germany}
\author{J. A.~Arregi}
\affiliation{CEITEC BUT, Brno University of Technology, 61200 Brno, Czech Republic}
\author{V.~Uhl\'{i}\v{r}}
\affiliation{CEITEC BUT, Brno University of Technology, 61200 Brno, Czech Republic}
\affiliation{Institute of Physical Engineering, Brno University of Technology, 61669 Brno, Czech Republic}
\author{M.~Bargheer}
\affiliation{Institut f\"ur Physik und Astronomie, Universit\"at Potsdam, 14476 Potsdam, Germany}
\affiliation{Helmholtz-Zentrum Berlin f\"ur Materialien und Energie GmbH, Wilhelm-Conrad-R\"ontgen Campus, BESSY II, 12489 Berlin, Germany}
\email{bargheer@uni-potsdam.de}

\date{\today}

\begin{abstract}
We use ultrafast x-ray diffraction to study non-equilibrium pathways of the phase transition in FeRh parametrized by the structural response. By increasing the pump-pulse duration beyond the electron-phonon coupling time, we suppress the electron-phonon non-equilibrium present upon femtosecond laser excitation but still photoexcite electrons to non-thermal states. Irrespective of the pump pulse duration, we find an optically induced nucleation of ferromagnetic domains on an $8\,\text{ps}$ timescale that starts as soon as the successively deposited energy surpasses the site-specific threshold energy. If in contrast, FeRh is only indirectly excited by diffusion of thermalized electrons from an opaque Pt cap layer, the ferromagnetic phase rises on a $50\,\text{ps}$ timescale. These findings unambiguously identify the photo-excitation of non-thermal electrons and not electron-phonon non-equilibria to enable the non-equilibrium pathway of the phase transition in FeRh. 
\end{abstract}

\maketitle

\section{Introduction}
The selective excitation of electrons or phonons by femtosecond laser pulses drives solids into non-thermal states\cite{webe2019,mald2020,muel2014,eich2017,weis2024,shok2024,soko2003} and phases\cite{jung2021,wall2013,john2023,koga2020} and opens transition routes and kinetics\cite{thie2023,wall2012,torr2021,john2024,davi2024} that are not accessible upon equilibrium heating. This includes coherent dynamics\cite{soko2003}, non-equilibria among different subsystems\cite{webe2019,mald2020} and non-thermal populations of certain degrees of freedom\cite{webe2019,mald2020,weis2024}. In addition to these non-thermal energy distributions, the laser-pulse can directly modify the electronic and magnetic structure via field-driven effects\cite{dewh2018,will2020,korf2024}.

In materials featuring first-order phase transitions, a fine interplay of spin, charge and lattice degrees of freedom\cite{jong2013, pole2016, wall2013,john2023,wegk2014} leads to an abrupt change of structural, electronic and magnetic properties. In this context, the non-equilibria introduced by femtosecond laser-excitation may open novel non-thermal routes of the phase transition through transient phases with different properties\cite{jung2021,john2023,koga2020,torr2021} where individually tracking the different degrees of freedom can provide unique insights into the microscopic processes and driving mechanisms of the phase transition\cite{wegk2014,wall2013,jong2013,jung2021}.

Since its discovery, the prototypical first-order magneto-structural antiferromagnetic-to-ferromagnetic (AFM-FM) phase transition of FeRh at $370\,\text{K}$ was extensively studied due to its potential for future applications\cite{lew2016,fen2019,fin2019,zhu2022}. Previous studies proposed a variety of different mechanisms driving the phase transition, for example expansion-induced sign change of the exchange constant\cite{kitt1960}, excitation of spin waves\cite{gu2005} and dominant FM exchange of the Fe moments mediated by an induced Rh moment\cite{grun2003, ju2004, sand2011, bark2015, pole2016, pres2021}. The latter was described as a competition between bilinear and higher-order four spin exchange terms in atomistic spin dynamics \cite{bark2015}, a combination of Heisenberg exchange of Fe and a Stoner model for Rh \cite{ju2004} and a modification of the Rh-Fe hybridization\cite{sand2011, pole2016, pres2021}. However, the microscopic pathway of the transition upon femtosecond laser-excitation is still under debate.

Previous time-resolved experiments individually tracked the evolution of the electronic, magnetic and lattice degrees of freedom. While photoemission spectroscopy identified a rapid modification of the electronic band structure within the first picosecond\cite{pres2021}, a macroscopic magnetization is only formed within hundreds of picoseconds via the alignment and coalescence of the nucleated FM domains\cite{berg2006, mari2012, li2022}. In contrast, directly probing the structural order parameter, i.e. the lattice expansion via ultrafast x-ray diffraction (UXRD) yields insights into the nucleation kinetics\cite{mari2012, qui2012, matt2023b, matt2023c}. Recently, we related the variety of reported rise times of the local FM phase to two distinct non- and near-equilibrium pathways of the phase transition\cite{matt2023b}. While FM domains nucleated near the surface on an $8\,\text{ps}$ timescale upon direct femtosecond photo-excitation, the ferromagnetic phase only emerges on a $50\,\text{ps}$ timescale upon heating above the transition temperature via heat diffusion to deeper regions. We demonstrated a drastic acceleration of the phase transition by tuning the optical absorption via lateral nano-structuring FeRh films favouring plasmonic absorption, which enables the non-equilibrium pathway in the entire volume\cite{matt2023c}. However, it remained unclear if this non-equilibrium pathway through the rapid modification of the electronic band structure is related to strongly heated electrons associated with an electron-phonon non-equilibrium or to field-effects and photo-excited non-thermal electrons as proposed previously\cite{pres2021}.

Here, we disentangle the role of photo-excited non-thermal electrons and electron-phonon non-equilibria for the laser-induced phase transition in FeRh by utilizing pump pulses with picosecond duration combined with femtosecond x-ray probe pulses accessing the evolution of the structural order parameter. Irrespective of the pump pulse duration between $0.06$ and $10.5\,\text{ps}$, we observe the FM phase to rise on an $8\,\text{ps}$ timescale where the final FM volume fraction is exclusively given by the total deposited energy and the transition starts as soon as the successively deposited energy overcomes the site-specific threshold of the first-order phase transition. These results indicate the decisive role of photo-excited non-thermal electrons for the dynamics of the laser-induced first-order phase transition in FeRh. To cross-check this conclusion, we additionally performed experiments on a thin FeRh layer that is only indirectly excited by heat transport via thermalized electrons generated in an optically opaque Pt capping layer that suppresses the direct optical excitation of FeRh. Under this condition, the FM phase rises only on a $50\,\text{ps}$ timescale irrespective of the excitation strength. 

Comparing the two experimental scenarios (i) direct photoexcitation of FeRh with long pump pulses that ensure electron-phonon equilibration and (ii) indirect excitation of FeRh via a faster, optically induced pulse of thermal energy in electrons and phonons, clearly shows that only the direct optical excitation of FeRh enables the ultrafast $8\,\text{ps}$ non-equilibrium pathway that seems to exploit the short non-thermalized situation of each photoexcited electron directly after its excitation. 
This finding makes the hypothesis of optical intersite spin and charge transfer to be responsible for the rapid change of the electronic band structure \cite{pres2021} and the subsequent emergence of the equilibrium FM phase via nucleation plausible.

\section{Experimental Details}
\begin{figure}[t!]
\centering
\includegraphics[width = \columnwidth]{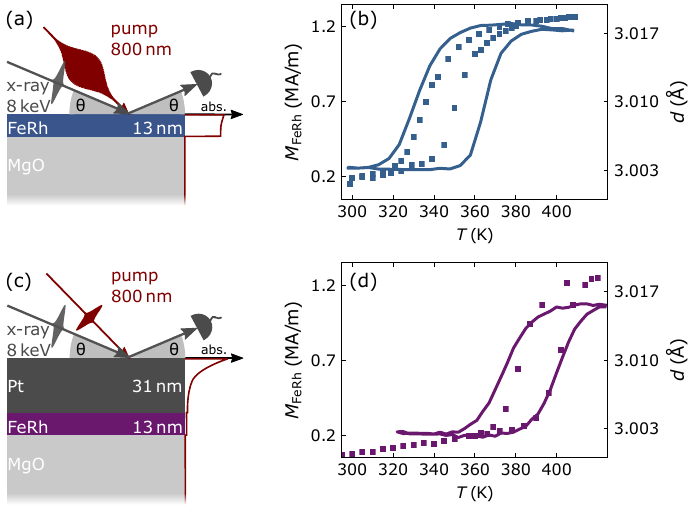}
\caption{\label{fig:fig_1_idea} \textbf{Static characterization of the FeRh films:} (a,c) Sketch of the FeRh samples, the optical absorption profiles and the UXRD experiment, mapping the reciprocal space via $\theta-2\theta$ scans. (b) Characterization of the AFM-FM phase transition in the bare FeRh film by the temperature-dependent magnetization (solid line) and the average out-of-plane lattice constant $d$ (symbols). (d) The same for the sample with Pt capping layer.}
\end{figure}

Figures~\ref{fig:fig_1_idea}(a) and (c) sketch the sample structures consisting of epitaxial $L_{\mathrm{FeRh}}=12.6\,\text{nm}$ thick FeRh(001) films grown by magnetron sputtering from an equiatomic FeRh target\cite{arre2020} on an MgO(001) substrate. The second sample is capped by a $30.8\,\text{nm}$ thick Pt(001) layer, which is much thicker than the inelastic mean free path of the electrons ($4\,\text{nm}$) \cite{zhuk2006} and the optical penetration depth ($\delta_\text{Pt}=9\,\text{nm}$) \cite{berg2020}. 
Previous experiments with variable Pt capping layer thickness reported a reduced heating of buried layers for thicknesses above $7\,\text{nm}$\cite{berg2020}. Therefore, the Pt capping layer ensures the indirect excitation of the buried FeRh layer by thermalized electrons. We used synchrotron radiation from the KMC-3 XPP endstation at BESSY II\cite{roes2021} to characterize the first order AFM-FM phase transition of the FeRh films via the concomitant change of the mean out-of-plane lattice constant $d$ (symbols in Figs.~\ref{fig:fig_1_idea}(b) and (d)). The hysteresis for this locally probed lattice constant is narrower than the global temperature-dependent magnetization $M_\text{FeRh}$ (solid line) determined by Vibrating sample magnetometry (VSM) using a QuantumDesign VersaLab magnetometer. The magnetization data indicate the presence of a residual FM phase of around $20\,\%$ originating from interface effects\cite{pres2016, fan2010, chen2017} consistent with the reduced out-of-plane expansion of $\eta^\text{thin}_\text{AFM-FM}=0.48\,\%$ compared to $\eta^\text{thick}_\text{AFM-FM}=0.6\,\%$ observed in thicker films\cite{matt2023b, matt2023c}.

Figures~\ref{fig:fig_1_idea}(a) and (c) sketch the performed UXRD experiments including the depth profile of the optically deposited energy. The sample structures are excited by $p$-polarized pump pulses with a central wavelength of $800\,\text{nm}$ that are incident under $40^\circ$ with respect to the sample normal. We probe the transient out-of-plane strain response of the FeRh layer via symmetric $\theta-2\theta$ scans \cite{schi2013a} around the FeRh(002) and Pt(002) Bragg peaks at a table-top laser-driven plasma x-ray source \cite{schi2012} providing $200\,\text{fs}$ hard x-ray pulses with a photon energy of approx. $8\,\text{keV}$. The Bragg peak position along the reciprocal space coordinate $q_z$ encodes the mean out-of-plane lattice constant $d$ of the FeRh films via $q_\text{z}=4\pi/d$. The lattice strain $\eta_\text{FeRh}=\Delta d/d_0$ is the relative change $\Delta d$ of the lattice constant with respect to its value $d_0$ before excitation. The bare FeRh film is excited by pump pulses of different durations between $0.06$ and $10.5\,\text{ps}$ realized by detuning a grating compressor. We independently determined the pump-probe overlap and cross-check the Gaussian profile of the pump-pulses by the strictly linear laser-induced response of a metal-insulator superlattice serving as reference sample\cite{schi2012}.

\section{Results and Discussion}
\subsection{Phase transition driven by picosecond pump pulses}
\begin{figure}[t!]
\centering
\includegraphics[width = \columnwidth]{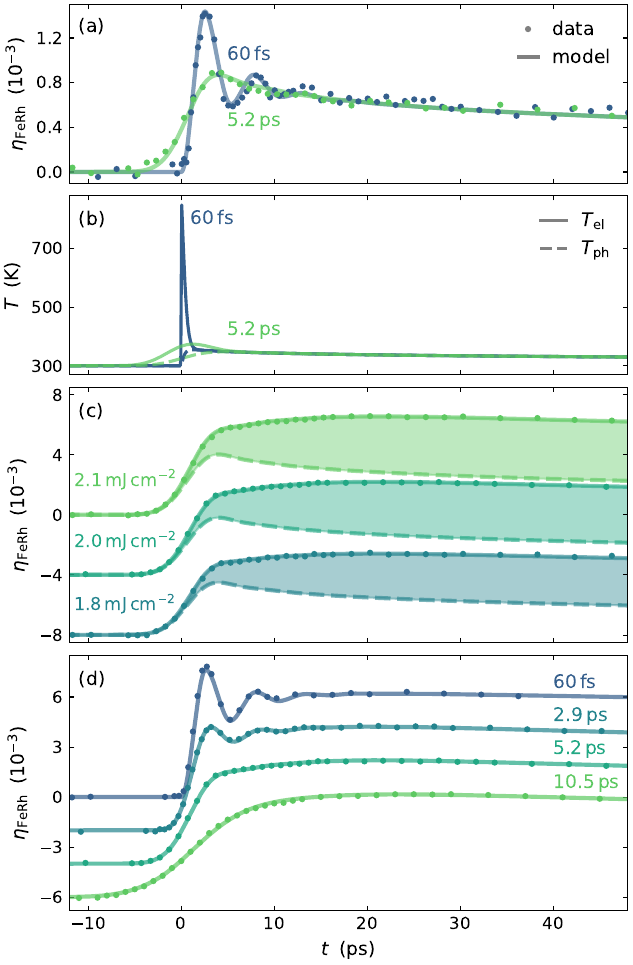}
\caption{\label{fig:fig_2_strain} \textbf{Transient strain response containing expansion from the phase transition:} (a) Thermoelastic strain response for a below-threshold excitation of $F_\text{bt}=0.4\,\text{mJ}\,\text{cm}^{-2}$ for $60\,\text{fs}$ and $5.2\,\text{ps}$-long pump pulses (symbols). The solid lines denote the strain response modeled by utilizing the \textsc{udkm1Dsim} library\cite{schi2021} and literature values for the thermo-elastic parameters. (b) The transient average electron and phonon temperatures in FeRh extracted from the strain modeling by applying a diffusive 2TM. (c) The strain response to various above-threshold excitations with $5.2\,\text{ps}$-long pump pulses additionally includes signatures of the driven AFM-FM phase transition highlighted by the coloured area representing the difference $\Delta \eta(t)$ between the measured strain and the fluence-scaled thermoelastic strain. (d) The strain response for a fixed excitation of $F_\text{at}=2.1\,\text{mJ}\,\text{cm}^{-2}$ and various pump pulse durations. The solid lines in (c) and (d) are the sum of strain due to the phase transition according to Eq.~\eqref{eq:eq_1_model} and the fluence-scaled sub-threshold strain response depicted in panel a). The results are off-set for clarity.}
\end{figure}

Figure~\ref{fig:fig_2_strain}(a) displays the laser-induced transient strain response of the bare FeRh film on MgO for a below-threshold excitation $F_\text{bt}=0.4\,\text{mJ}\,\text{cm}^{-2}$ with pump pulses of $60\,\text{fs}$ and $5.2\,\text{ps}$ duration that do not drive the AFM-FM phase transition. Thus, the strain response upon femtosecond laser excitation is the superposition of only two contributions: (i) A quasi-static expansion originating from heating the electrons and phonons of FeRh and (ii) propagating strain pulses driven by the rapidly rising unbalanced stress at the surface and the substrate interface upon femtosecond laser excitation\cite{matt2023a}. The launched strain pulse propagates with sound velocity $v_s$ and is reflected at the surface and partially transmitted into the substrate. This leads to a decaying oscillation\cite{matt2023a} with a period of $2L_\text{FeRh}/v_s$ that is superimposed with a decreasing quasi-static expansion due to heat transport into the substrate. The $5.2\,\text{ps}$ pump pulse slowly heats FeRh. This drives a stress on the lattice which rises slower than its relaxation by lattice expansion with sound velocity. Therefore, pump pulses with durations significantly exceeding $2L_\text{FeRh}/v_s$ disable the coherent excitation of picosecond strain pulses and thus only the slowly varying quasi-static expansion contribution to the strain response remains. Since the strain pulse launched by femtosecond laser excitation completely enters the MgO substrate within $20\,\text{ps}$, the subsequent evolution of the strain response is independent of the pump pulse duration.

The solid lines in Fig.~\ref{fig:fig_2_strain}(a) denote our strain model utilizing the modular \textsc{Python} library \textsc{udkm1Dsim} \cite{schi2021} and literature values for the thermoelastic parameters given in Tab.~\ref{tab:tab_1_params}. We apply a diffusive two-temperature model (2TM) to describe the spatio-temporal distribution of the optically deposited energy among electrons and phonons. The transient energy density stored in those degrees of freedom is linearly related to the stress on the lattice by subsystem-specific Gr\"uneisen parameters. Their superposition drives the lattice response according to the linear one-dimensional elastic wave equation. We find excellent agreement of our model with the experiment, if we assume a Gaussian shape of the pump pulses and consider their experimentally determined durations. We emphasize that the green line, which represents the modeling for a $5.2\,\text{ps}$ pump pulse is indistinguishable from the convolution of the strain model for $60\,\text{fs}$ (blue line) with a Gaussian of $5.2\,\text{ps}$ full-width half-maximum. This is consistent with the strictly linear thermoelastic strain response expected for metals as a function of the fluence\cite{matt2023a}. Figure~\ref{fig:fig_2_strain}(b) displays the respective transient electron and phonon temperatures. While the electrons and phonons experience a pronounced non-equilibrium upon femtosecond laser-excitation, the $5.2\,\text{ps}$-long pump pulse suppresses this non-equilibrium by drastically reducing the maximum electron temperature because a considerable amount of energy is already dissipated to the phonons during the optical absorption.
\begin{table}[t!]
\centering
\begin{ruledtabular}
\begin{tabular}{l c c c}
 & Pt & FeRh & MgO\\
 \hline
$\delta_\mathrm{p}$ (nm) & 9\cite{berg2020} & 15\cite{matt2023c} & inf\\
$\gamma\textsuperscript{S}$ (mJ\,cm\textsuperscript{-3}\,K\textsuperscript{-2})& 0.73\cite{hohl2000} & 0.16\cite{tu1969}& -\\
$C\textsubscript{ph}$ (J\,cm\textsuperscript{-3}\,K\textsuperscript{-1}) & 2.85\cite{shay2016} & 3.48\cite{rich1973} & 3.32 \cite{barr1959}\\
$\kappa_\text{el}^0$ (W\,m\textsuperscript{-1}\,K\textsuperscript{-1}) & 66\cite{dugg1970} & 45 & - \\
$\kappa\textsubscript{ph}$ (W\,m\textsuperscript{-1}\,K\textsuperscript{-1}) &5.0\cite{dugg1970} & 5.0 & 50 \cite{slif1998}\\
$g$ (PW\,m\textsuperscript{-3}\,K\textsuperscript{-1})& 375\cite{zahn2021} & 900\cite{gunt2014} & -\\
$\rho$ (g\,cm\textsuperscript{-3}) & 21.45 & 9.93 & 3.58\\
$v\textsubscript{S}$ (nm\,ps\textsuperscript{-1})& 4.0\cite{farl1966} & 5.0 & 9.1 \cite{dura1936}\\
$\Gamma$\textsubscript{el} & 1.2\cite{jare2024} & 1.4 & - \\
$\Gamma$\textsubscript{ph} & 2.6\cite{nix1942} & 1.7\cite{ibar1994} & 1.7 \cite{whit1966}\\
\end{tabular}
\end{ruledtabular}
\caption{Literature values for the physical parameters of the strain model. The optical penetration depth $\delta_\text{p}$, the Sommerfeld constant $\gamma^\text{S}$, the specific heat of the phonons $C_\text{ph}$, the electron-phonon coupling constant $g$, and the electron $\kappa_e^0$ and phonon $\kappa\textsubscript{ph}$ heat conductivity determine the spatio-temporal energy distribution upon laser-excitation in the framework of a diffusive two-temperature model\cite{matt2023a}. The subsystem-specific Gr\"uneisen parameters $\Gamma_\text{el}$ and $\Gamma_\text{ph}$ linearly relate the spatio-temporal energy density to an elastic stress on the lattice driving a quasi-static expansion and strain pulses propagating with sound velocity $v_\text{S}$ according to the elastic wave equation\cite{matt2023a}.We additionally reduced the phonon conductivity of the first MgO unit cell to  $0.5\,\text{W}\,\text{m}^{-1}\,\text{K}^{-1}$ and the electron conductivity of the last Pt unit cell to $0.5\,\text{W}\,\text{m}^{-1}\,\text{K}^{-1}$ to mimic an interface resistance\cite{matt2023a}.}
\label{tab:tab_1_params}
\end{table}

In the following, we utilize this calibration of the thermoelastic strain $\eta_\text{FeRh}^\text{bt}$ to extract the signatures of the AFM-FM phase transition from the strain response $\eta_\text{FeRh}^\text{at}$ to above-threshold excitations in Figs.~\ref{fig:fig_2_strain}(c) and (d). Figure~\ref{fig:fig_2_strain}(c) exemplarily displays the signature of the laser-induced phase transition to the strain response for various above-threshold excitations by pump pulses of $5.2\,\text{ps}$ duration. The dashed lines denote the modeled thermoelastic strain contributions for $F_\text{bt}=0.4\,\text{mJ}\,\text{cm}^{-2}$ (green solid line in (a)) scaled to the fluences $F_\text{at}$ in the experiment. The difference to the actual measurement $\Delta \eta(t)$ highlighted by coloured areas is related to an additional expansion that parametrizes the phase transition\cite{matt2023b}. To determine the absolute value of the transient FM volume fraction, we consider the expansion across the phase transition in thermal equilibrium $\eta^\text{thick}_\text{AFM-FM}$ and the latent heat $C_\text{AFM-FM}$ of the first-order phase transition: The energy $Q_\text{AFM-FM}=\int C_\text{AFM-FM} dT$ required for transforming FeRh into the FM phase reduces the local temperature\cite{ahn2022} by $\Delta T=Q_\text{AFM-FM}/C_\text{ph}$, which reduces the quasi-static expansion of FeRh by $\eta_\text{AFM}=\alpha_\text{AFM}\Delta T$ with the expansion coefficient $\alpha_\text{AFM}$ in the AFM phase. This relates $\Delta \eta(t)=\eta^\text{at}_\text{FeRh}-F_\text{at}/F_\text{bt} \cdot\eta^\text{bt}_\text{FeRh}$ to the laser-induced FM volume fraction $\Delta V_\text{FM}$ via:
\begin{equation}
    \Delta \eta(t)= (\eta^\text{thick}_\text{AFM-FM}-\eta_\text{AFM})\Delta V_\text{FM}(t)\,,
    \label{eq:eq_0_VFM}
\end{equation}
where the total FM fraction $V_\text{FM}(t)=\Delta V_\text{FM}(t)+V^0_\text{FM}$ additionally includes the residual FM phase $V^0_\text{FM}=0.2$ present before the laser excitation as characterized in Fig.~\ref{fig:fig_1_idea}(b). Figure~\ref{fig:fig_2_strain}(d) displays the strain response for systematically varied pump pulse durations and an excitation of $2.1\,\text{mJ}\,\text{cm}^{-2}$ driving the phase transition as identified by the fluence dependence in panel (c). With increasing pump pulse duration the oscillations in the strain response originating from coherently driven picosecond strain pulses are successively suppressed and completely vanish for $5.2\,\text{ps}$-long pump pulses. In contrast to the coherent dynamics, the expansion of FeRh beyond $20\,\text{ps}$ is identical for all pump pulse durations. This indicates that also the final FM volume fraction $V_\text{FM}^*$ is independent of the pump pulse duration.

\begin{figure}[t!]
\centering
\includegraphics[width = \columnwidth]{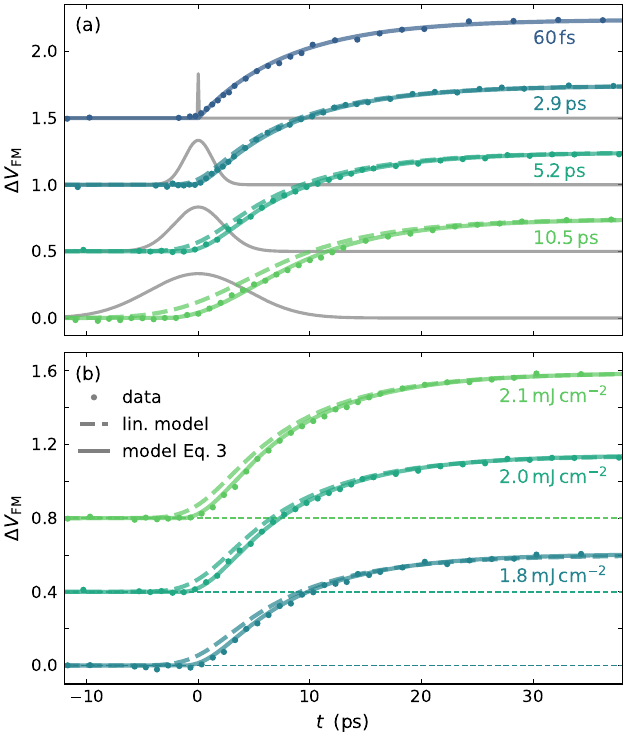}
\caption{\label{fig:fig_3_FM} \textbf{Laser-induced phase transition:} (a) Laser-induced rise of the FM volume fraction $\Delta V_\text{FM}$ for different pump-pulse durations (assumed pulse profile indicated by grey solid lines) and a fluence of $2.1\,\text{mJ}\,\text{cm}^{-2}$. (b) Laser-induced rise of $\Delta V_\text{FM}$ for various above-threshold excitations and $5.2\,\text{ps}$ pump-pulse duration. In both panels, the dashed lines represent the FM volume fraction according to Eq.~\eqref{eq:eq_0_model_old}, i.e. for ultrashort pulse excitation, convoluted with a Gaussian representing the duration of the respective pump-pulse. The data rise faster but start delayed compared to the dashed lines, which correspond to a linear response via the convolution. The solid lines show the appropriate model for the FM volume fraction according to Eq.~\eqref{eq:eq_1_model}, which accounts for the non-linear threshold behavior. The plots are off-set for clarity.}
\end{figure}

Figure~\ref{fig:fig_3_FM} displays the extracted laser-induced FM volume fraction $\Delta V_\text{FM}(t)$ that parameterizes the optically driven phase transition. For $60\,\text{fs}$ pump pulses, $\Delta V_\text{FM}(t)$ in Fig.~\ref{fig:fig_3_FM}(a) is very well reproduced by a single exponential rise associated with the nucleation of domains\cite{mari2012} on an $\tau=8\,\text{ps}$ timescale\cite{matt2023b} by:
\begin{align}
    \Delta V_\text{FM}(t) &= \mathcal{H}(t) \Delta V_\text{FM}^* \cdot \left( 1-e^{-t/\tau} \right) \; ,
\label{eq:eq_0_model_old}
\end{align}
with the Heaviside function $\mathcal{H}(t)$ and the final FM volume fraction increase $\Delta V_\text{FM}^*$ that depends on the fluence with $\Delta V_\text{FM}^*>0$ if $F>F_\text{th}=0.6\,\text{mJ}\,\text{cm}^{-2}$ and $\Delta V_\text{FM}^*=0.8$ if $F=2.1\,\text{mJ}\,\text{cm}^{-2}$ as calibrated previously\cite{matt2023b}. This fluence dependence upon femtosecond laser excitation including $\Delta V_\text{FM}^*=0.6$ for $1.8\,\text{mJ}\,\text{cm}^{-2}$ is perfectly reproduced by the fluence-dependent $\Delta V_\text{FM}(t)$ upon excitation by $5.2\,\text{ps}$ pump pulses in Fig.~\ref{fig:fig_3_FM}(b). This agreement demonstrates an identical threshold of the first-order phase transition irrespective of the pump pulse duration. We emphasize that the non-linear response of the first-order phase transition is seen in the fact that a tiny increase of the fluence from $1.8$ to $2.1\,\text{mJ}\,\text{cm}^{-2}$ changes $\Delta V_\text{FM}$ from $60\,\%$ to a complete phase transition, which corresponds to $\Delta V_\text{FM}= 80\,\%$, because of the initial residual ferromagnetic volume fraction.

The dashed lines in Fig.~\ref{fig:fig_3_FM} represent $\Delta V_\text{FM}(t)$ from Eq.~\eqref{eq:eq_0_model_old} convoluted with a Gaussian representing the pump-pulse duration in the experiment. This approach assuming a linear response to the optical excitation was used to fit the laser-induced demagnetization in Ni associated with a second order phase transition\cite{fogn2015}. However, for the first-order phase transition in FeRh this approach clearly deviates from the measurement for long pump pulses. With increasing pump pulse duration this deviation within the first picoseconds becomes more pronounced (see Fig.~\ref{fig:fig_3_FM}(a)). Interestingly, the data rise faster than the convolution although the rise starts later, which we attribute to the non-linearity associated with the threshold for the phase transition.

As an improved model of $\Delta V_\text{FM}(t)$ for long pump-pulses, we explicitly consider the successively deposited energy that leads to the unlocking of the AFM-FM phase transition in an increasing volume fraction of the film during the pump-pulse. This explicit treatment of the threshold character of first-order phase transitions extends Eq.~\eqref{eq:eq_0_model_old} in the case of picosecond pump-pulses to
\begin{align}
    \Delta V_\text{FM}(t) &= \int \mathcal{H}(t-t^\prime) V_\text{FM}^*(t^\prime) \cdot \left( 1-e^{-(t-t^\prime)/\tau} \right) dt^\prime \; .
\label{eq:eq_1_model}
\end{align}
Here, the FM volume fraction $V_\text{FM}^*(t^\prime)$ is unlocked at delay $t^\prime$ by the increase of the deposited energy and subsequently rises on the $\tau = 8\,\text{ps}$ nucleation timescale. This transiently unlocked phase transition adds to the already present FM phase driven at delays $t<t^\prime$. The Heaviside function $\mathcal{H}(t-t^\prime)$ ensures a start of the phase transition at $t^\prime$. To model the transient FM volume fraction we assume a linear increase of $V_\text{FM}^*$ from $0$ to $0.8$ between the fluences $F_\text{th}=0.6\,\text{mJ}\,\text{cm}^{-2}$ and $2.1\,\text{mJ}\,\text{cm}^{-2}$, which corresponds to a full phase transition\cite{matt2023b}.

Under this assumption, Eq.~\eqref{eq:eq_1_model} yields excellent agreement (solid lines) with the experimentally determined transients $\Delta V_\text{FM}(t)$ in Figs.~\ref{fig:fig_3_FM}(a) and (b) for various pump-pulse durations and above-threshold fluences. 
The model also agrees very well with the measured dependence of the transient strain on the fluence and pulse duration in Fig.~\ref{fig:fig_2_strain}(c) and (d). The solid lines represent the sum $\eta^\text{bt}_\text{FeRh}(t)=F_\text{at}/F_\text{bt} \cdot\eta^\text{bt}_\text{FeRh}(t)+\Delta \eta (t)$, i.e. the fluence-scaled thermoelastic below threshold strain plus the strain contribution $\Delta \eta (t)$ determined in Fig.~\ref{fig:fig_3_FM} using Eq.~\eqref{eq:eq_1_model}. Our model accounts for the threshold of the first-order phase transition. Therefore it can reproduce both the delayed start of the nucleation of FM domains relative to the beginning of the pump pulse for longer pump-pulses and decreasing fluence. Furthermore, it demonstrates that the nucleation of FM domains proceeds on the intrinsic $8\,\text{ps}$ timescale upon laser excitation even for $10.5\,\text{ps}$ long pump pulses that suppress any electron-phonon non-equilibrium.

\subsection{Phase transition upon near-equilibrium heating}
These results obtained in a homogeneously excited $12\,\text{nm}$ FeRh film show that the direct optical excitation always triggers the phase transition within $8\,\text{ps}$. On the other hand, our recent experiment on thicker inhomogeneously excited FeRh films show that longer timescales may apply in the deeper parts of the film, which are not directly optically excited \cite{matt2023b}. 
In order to cross-check this insight in a more directly comparable FeRh thin film, we performed an UXRD experiment on a similar $13\,\text{nm}$ thin FeRh film buried below an optically opaque Pt layer (see Fig.~\ref{fig:fig_1_idea}(c)) that suppresses the direct optical excitation of FeRh. It is then only excited by heat transport via thermalized electrons.

\begin{figure}[t!]
\centering
\includegraphics[width = \columnwidth]{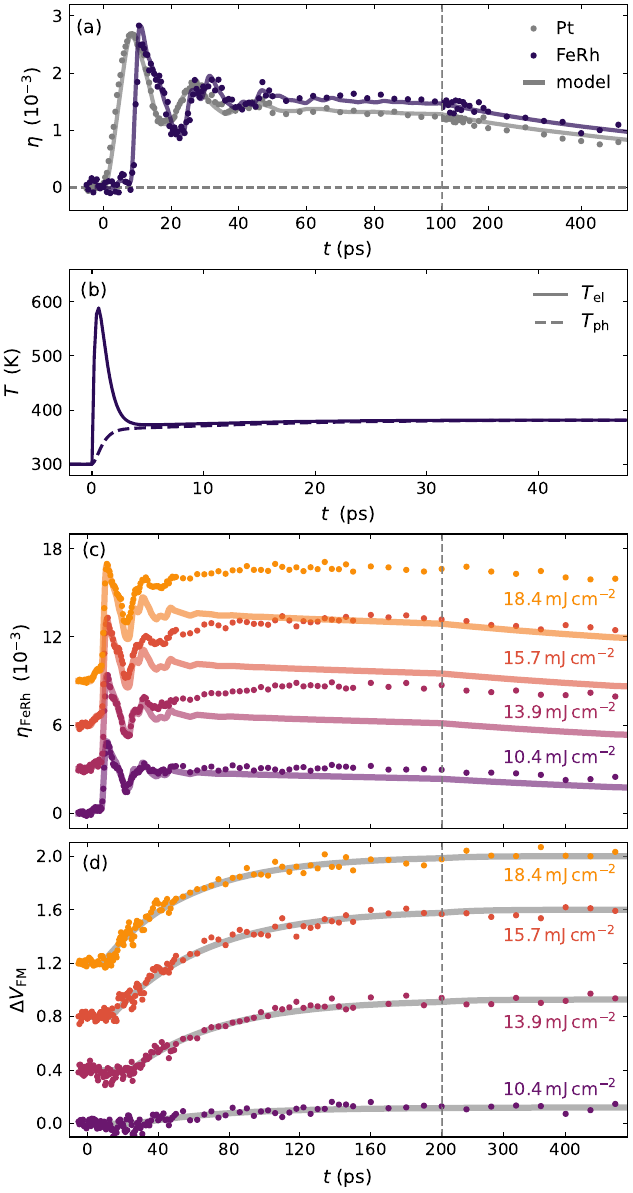}
\caption{\label{fig:fig_4_Pt} \textbf{Dynamics of phase transition in case of suppressed direct optical excitation:} (a) The strain response of the Pt and the FeRh layer to a below-threshold excitation of $F_\text{bt}=5.8\,\text{mJ}\,\text{cm}^{-2}$. The solid lines represent the modeled strain utilizing the \textsc{Python} library \textsc{udkm1Dsim} and the thermoelastic parameters in Tab.~\ref{tab:tab_1_params}. (b) The average electron and phonon temperatures in FeRh extracted from the diffusive 2TM of the strain modeling for the same fluence. (c) The strain response of FeRh to above-threshold excitations (symbols). The additional expansion compared to the modeled thermoelastic strain response (solid lines) is a clear signature of the laser-induced phase transition. (d) Extracted transient change of the FM volume fraction $\Delta V_\text{FM}$ according to Eq.~\eqref{eq:eq_0_VFM} (symbols). The gray solid lines denote a single exponential fit to the experimental results with an average rise time of $50.7 \pm 1.9\,\text{ps}$. The results in (c) and (d) are off-set for clarity.}
\end{figure}

As in section A for the sample without Pt cap, we measured and modeled the strain response of the FeRh layer to a below-threshold excitation. To make our modeling more reliable, we additionally measured and modeled the strain response of the Pt capping layer. Figure~\ref{fig:fig_4_Pt}(a) shows an excellent agreement of our model with the transient strain response of both Pt and FeRh with a single set of parameters (see Tab.~\ref{tab:tab_1_params}). Pronounced features of the dynamics are the direct expansion of Pt and the clear $8.5\,\text{ps}$ delay of the sharp onset of the FeRh expansion, that precisely matches the time that the longitudinal acoustic strain pulse takes to travel through Pt. Figure~\ref{fig:fig_4_Pt}(b) displays the associated average electron and phonon temperatures in FeRh. We find a rapid heating of the electrons within the first picosecond via heat diffusion from Pt. However, even after the electron-phonon equilibration  within $5\,\text{ps}$, energy is slowly transported to FeRh as indicated by the slowly rising electron and phonon temperature converging to the maximum within $35\,\text{ps}$. This indirect excitation leads to a delay of the FeRh expansion with respect to the Pt capping layer (see Fig.~\ref{fig:fig_4_Pt}(a)). The slow equilibration of the layers results in the slightly increasing quasi-static expansion of FeRh between $20$ and $50\,\text{ps}$ while the quasi-static expansion of Pt already decays.

With this model for the thermoelastic strain contribution at hand, we extract the additional expansion contribution associated with the driven AFM-FM phase transition $\Delta \eta(t)$ from the strain response to above threshold excitations $\eta_\text{at}(t)$ shown in Fig.~\ref{fig:fig_4_Pt}(c). The solid lines denote the thermoelastic strain contribution scaled to the fluence and the deviation to the actual measurement directly identifies the laser-induced phase transition. By applying Eq.~\eqref{eq:eq_0_VFM}, we extract the corresponding change of the FM volume fraction $\Delta V_\text{FM}$. The grey solid lines denote a fit of the transient FM volume fraction by a single exponential with a variable timescale and a variable delay $t_T$ at which the FM phase transition starts. Irrespective of the excitation strength, we find a $\tau_\text{eq}=50.7 \pm 1.9\,\text{ps}$ timescale for the rise of the FM phase according to $1-e^{-(t-t_T)/\tau_\text{eq}}$, with an offset time $t_T$.

All data in  Fig.~\ref{fig:fig_4_Pt}(d) can be fitted with the same $\tau_\text{eq}$, with the clear trend that higher fluences lead to an earlier start $t_T$ of the nucleation and a larger converted volume fraction $\Delta V_\text{FM}^*$. However, the precise timing and amplitude of the emerging FM phase is not described by our straight-forward thermodynamic model that yields the excellent agreement with the sub-threshold strain in Fig.~\ref{fig:fig_4_Pt}(a). Although the fluence scaling reproduces the acoustic strain waves in the first $40\,\text{ps}$ in Fig.~\ref{fig:fig_4_Pt}(c), the modeled phonon temperature in FeRh for indirect excitation through Pt at $10.4\,\text{mJ}\,\text{cm}^{-2}$ exceeds $440\,\text{K}$. According to the characterisation in thermal equilibrium in Fig.~\ref{fig:fig_1_idea}(d), this should drive a full phase transition in the entire film and not only $15\,\%$. Even if we consider the energy consumed by the latent heat of the phase transition and the enhanced transition temperature due to the tetragonal distortion of the unit cell by $4\,\text{K}$ per $0.1\,\%$ out-of-plane expansion\cite{fin2019}, we would expect a complete phase transition from the behaviour in thermal equilibrium. 
Furthermore, the delay $t_T$ at which the phase transition starts, is not simply given by the time at which the phonon temperature exceeds the equilibrium phase transition temperature. It seems that for higher fluences ($15.7$ and $18.4\,\text{mJ/cm}^2$) the nucleation starts when the expansion wave launched at the Pt surface reaches the FeRh layer at about $8.5\,\text{ps}$. For lower fluences, the nucleation seems to start only when the second expansion wave after the sharp strain minimum at $25\,\text{ps}$ (Fig.~\ref{fig:fig_4_Pt}(c)) enters FeRh. We suggest that under near-equilibrium heating the FM phase emerges similar to experiments in thermal equilibrium that reported a nucleation of FM domains as columns through the entire film and a subsequent in-plane domain growth\cite{gate2017}. Assuming that the columnar grains composing our epitaxial film \cite{arre2020} act as a fundamental block for the phase transition, the top part of the FeRh film can not transition until the bottom part is sufficiently heated too. In a simple picture this means that the FM phase can only emerge when the phonon temperature exceeds the transition temperature through the entire thickness of the FeRh film. Our thermodynamic model shows that for these fluences the phonon temperature at the backside of the layer is still below the transition temperature at $8.5\,\text{ps}$, which delays the emergence of the FM phase until the reflection from the surface enters FeRh at $25\,\text{ps}$ as an expansive strain wave. This indicates that the nucleation of the FM phase in a superheated AFM state of FeRh is started by an impulsive expansion, similar to the nucleation of ice in a supercooled liquid \cite{debe2001}. Despite the fact that the coherent strain pulse must be faster than the diffusion of heat via phonons, the first strain wave can indeed impinge on a superheated FeRh as the electrons rapidly transport energy from Pt into FeRh according to the 2TM (Fig.~\ref{fig:fig_4_Pt}(b)). 

The quantitative understanding of the observed volume fractions in the near-equilibrium pathway of the phase transition in FeRh may require a model beyond a simple 2TM approximation that accounts for the energy in the relevant degrees of freedom. However, the confirmation of the $50\,\text{ps}$ timescale associated with the near-equilibrium pathway of the phase transition in FeRh\cite{matt2023b} by burying FeRh below an optically opaque Pt layer proves that indeed the photo-excitation of electrons to non-thermal states is required to enable the faster non-equilibrium pathway of the phase transition in FeRh. This matches the hypothesis that optically induced intersite spin and change transfer lead to a rapid modification of the electronic band structure\cite{pres2021} that is followed by establishing the equilibrium FM phase via the nucleation of FM domains on the $8\,\text{ps}$ timescale.

\section{Conclusion}
In summary, we studied the laser-induced magnetostructural AFM-FM phase transition in FeRh upon direct photo-excitation with variable pump pulse duration and upon indirect excitation via heat transport by adding an opaque Pt capping layer. Irrespective of the pump pulse duration up to $10.5\,\text{ps}$, which exceeds the electron-phonon coupling time by far, we find the FM phase to nucleate on the $8\,\text{ps}$ timescale associated with the non-equilibrium pathway of the phase transition. In contrast, the excitation of FeRh with thermalized electrons optically generated in a Pt cap layer leads to the rise of the FM phase on the $50\,\text{ps}$ timescale associated with the near-equilibrium pathway of the phase transition. These results reveal the crucial role of direct photo-excitation of non-thermal electrons for the ultrafast pathway of the phase transition in FeRh and the negligible relevance of strongly elevated electron temperatures. This supports the previous theoretical prediction of optically induced charge and spin transfer among Fe- and Rh-sites to be responsible for the rapid modification of the electronic band structure that is followed by the establishing of the equilibrium FM phase via nucleation\cite{pres2021}. Our experiments illustrate that picosecond pump pulses are generally useful to identify the non-linear threshold behaviour of first-order phase transitions and to disentangle the role of non-thermal and strongly heated but thermalized electrons for (phase) transition dynamics.

\section*{Acknowledgements}
We acknowledge the DFG for financial support via Project-No.\ 328545488 – TRR 227, project A10 and the BMBF for funding via 05K22IP1. Access to the CEITEC Nano Research Infrastructure was supported by the Ministry of Education, Youth and Sports (MEYS) of the Czech Republic under the project CzechNanoLab (LM2023051). Beamtimes at the KMC-3 XPP endstation of the
synchrotron radiation facility BESSY II at the Helmholtz Zentrum
Berlin were required for thorough sample characterization.


\end{document}